\documentstyle[12pt,epsf]{article}
\newcommand{\fc}{\mbox{$ f_c \; $}}

\newcommand{\tauh}{\mbox{$ {\tau}_{\scriptstyle hom} \; $}}
\newcommand{\taueff}{\mbox{$ {\tau}_{\scriptstyle eff} \; $}}
\newcommand{\weff}{\mbox{$ {\omega}_{\scriptstyle eff} \; $}}
\newcommand{\Reff}{\mbox{$ {R}_{\scriptstyle eff} \; $}}

\begin{document}

\begin{center}

{\LARGE \bf Radiative Transfer in Clumpy and Fractal Media }

\bigskip                
\bigskip                

{\large Frank V\'arosi $^{* \dagger}$ \, and \, Eli Dwek}

\bigskip                

{\small \sl  Infrared Astrophysics Branch, Code 685, \\
	NASA/Goddard Space Flight Center,
	Greenbelt, MD 20771 \\
	$^{\dagger}$ Raytheon ITSS Corp., Lanham, MD 20706 }

\bigskip                

{\small Appeared in The Ultraviolet Universe at Low and High Redshift, \\
eds. W. Waller et al. (New York: AIP), p.370, \\
proceedings of the conference held at University of Maryland in 1997}

\bigskip                

\end{center}

\pagestyle{empty}

\begin{abstract}
A Monte Carlo model of radiative transfer in multi-phase dusty media is
applied to the situation of stars and clumpy dust in a sphere or a disk.
The distribution of escaping and absorbed photons are shown for
various filling factors and densities.  Analytical methods of
approximating the escaping fraction of radiation, based on the
Mega-Grains approach \cite{hobpad93}, are discussed.  Comparison with
the Monte Carlo results shows that the escape probability formulae
provide a reasonable approximation of the escaping/absorbed fractions,
for a wide range of parameters characterizing a clumpy dusty medium.  A
possibly more realistic model of the interstellar medium is one in
which clouds have a self-similar hierarchical structure of denser and
denser clumps within clumps \cite{elm97}, resulting in a fractal
distribution of gas and dust.  Monte Carlo simulations of radiative
transfer in such multi-phase fractal media are compared with the
two-phase clumpy case.
\end{abstract}

\section{Introduction}

Radiative transfer plays an important role in the evolution of the
spectral appearance of galaxies, and at wavelengths longer than the
Lyman limit, scattering by dust in the interstellar medium (ISM)
is the major factor. The ISM is known to be composed of at least three phases:
diffuse clouds, dense molecular clouds,
and a low density inter-cloud medium (ICM).
Most likely the ISM has a spectrum of densities and temperatures,
as proposed by theoretical models \cite{norman96} \cite{elm97},
with correlated multi-scale spatial structure,
as evidenced by sky surveys such as IRAS and H I radio surveys.
The transfer of radiation becomes complicated in such an inhomogeneous medium,
however in most cases the effective optical depth is less than that of
the homogeneous medium with equal mass of dust, allowing relatively more
photons to escape.

\pagestyle{myheadings}
\markright{F.V\'arosi \& E.Dwek: Radiative Transfer in Clumpy and Fractal Media}

The simplest model of an inhomogeneous medium is two phases (densities):
dense clumps of dust in a less dense ICM.
Radiative transfer through such a clumpy plane-parallel medium was investigated
by Boiss\'e \cite{boisse90},
and then by Hobson \& Scheuer \cite{hs93} for two and three phase media.
Their results for a three phase medium were found
to be significantly different than the two phase case.
Recently Witt \& Gordon \cite{wittgor96}
performed Monte Carlo simulations of radiative transfer
from a central source in a two phase clumpy medium within a sphere.
All these investigations verify the expectation that the medium becomes
more transparent as the degree of clumpiness is increased.

\section{Monte Carlo Simulations}

To further explore these effects we have developed a Monte Carlo code
for simulating radiative transfer
with multiple scattering in an inhomogeneous dusty medium
\cite{witt77} \cite{LuxKob95}.
The geometry and density of the medium can be specified
by a continuous functional $\rho(x,y,z)$, or on a three dimensional grid.
For each wavelength, the number of photons absorbed by the dust in each element
of the 3D grid is saved, allowing computation of the dust temperatures and 
resulting infrared emission spectrum.
The grid resolution is limited only by the available computer memory:
increasing the number of grid elements does not affect the computation time.
This is achieved by employing the Monte Carlo method of
imaginary/real scatterings and rejections \cite{LuxKob95}
in selecting the random distances each photon travels between interactions,
instead of numerical integration across volume elements of the grid.  
Our Monte Carlo simulations
agree exactly with the radiative transfer results of
Witt \& Gordon for the situation they considered,
that of cubic clumps on a body centered cubic percolation lattice.
However, we find that randomly located spherical clumps
create a more natural two phase medium,
and the radiative transfer properties can then be approximated
by the Mega-Grains approach of Hobson \& Padman \cite{hobpad93}.

Since the ISM has a wide spectrum of densities controlled by both
compressible turbulence and gravitational collapse \cite{norman96} \cite{elm97},
the structure of gas and dust clouds is more likely
a self-similar hierarchy of denser clumps within clumps.
A fractal distribution of matter has exactly such properties.
We construct fractal cloud models
using a modification of the algorithm described by Elmegreen \cite{elm97},
then create density maps on a 3D grid,
considering everything outside of the fractal cloud to be the low density ICM.
The maximum density contrast is determined by the resolution
of the density map grid. Any subset which is much larger than the
resolution limit is also a fractal cloud, by the self-similar construction.

%-------------------------------------------------------------------- 

\begin{figure}
\label{absorption_maps}                               
  \epsfysize=6.3in
  \centerline{\vbox{\epsfbox{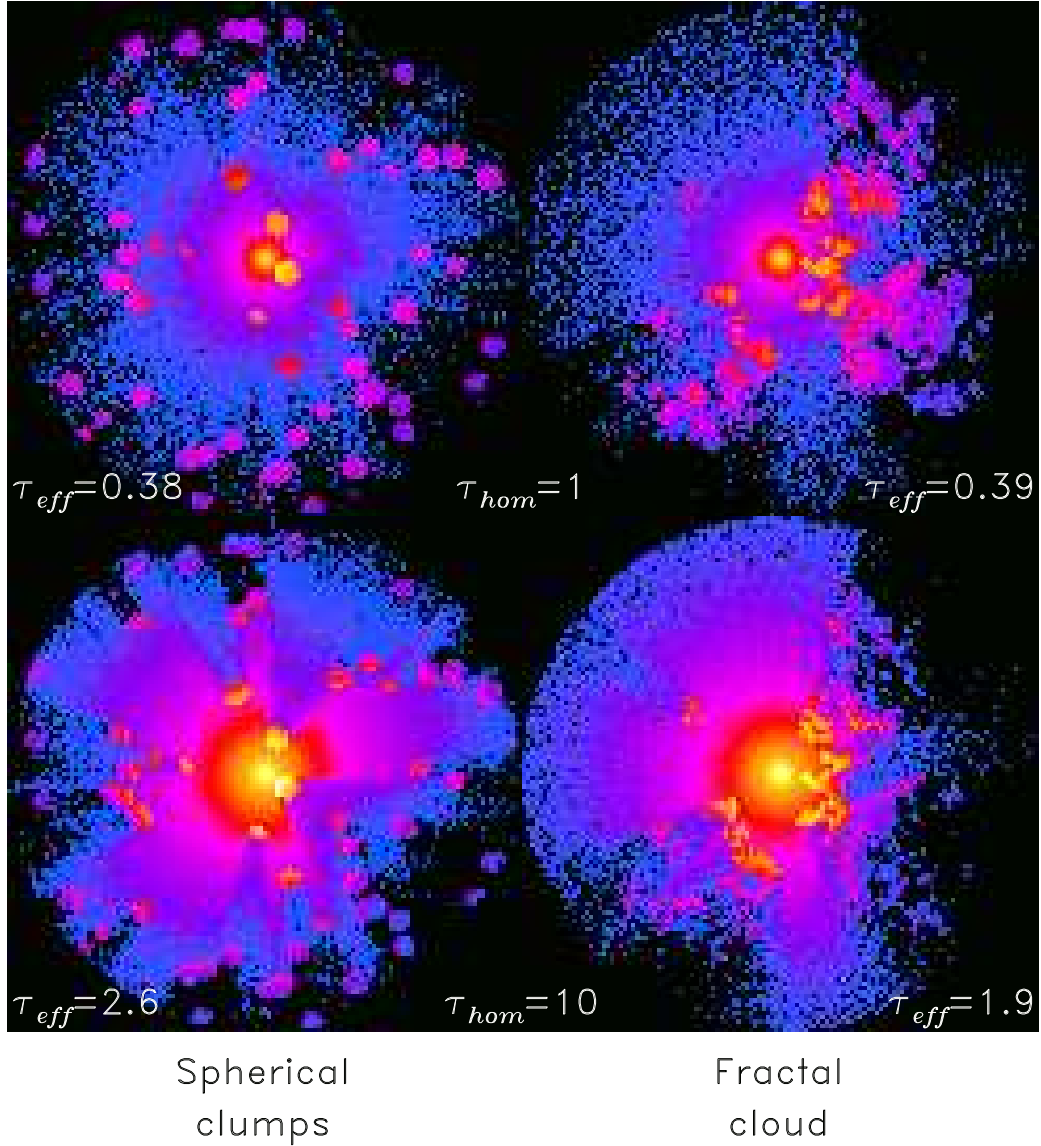}}}
\caption{
Maps of the photons from a central source that are absorbed
by dust in a 2D slice through the center of two types of inhomogeneous media,
as simulated by Monte Carlo methods.
For each type of media (each column), simulated maps are shown at
two values of the equivalent homogeneous optical depth \tauh,
which is the radial optical depth of absorption and scattering that would result
if the dust was distributed uniformly in the sphere.
The volume filling factor of the clumps and fractal cloud is $\fc=0.1$.
Blue coloring indicates the minimum absorption,
red indicates more absorption, yellow is maximum,
and the logarithmic scaling over five orders of magnitude
is the same for each map.
See text for more information.
}
\end{figure}

%-----------------------------------------------------------------------------
 
Figure~1 shows an array of images,
each one being a map of the photons from a central source that are absorbed
by dust in a 2D slice through the center of two types of inhomogeneous media.
Blue coloring indicates the minimum absorption,
red indicates more absorption, yellow is maximum,
and the logarithmic scaling over five orders of magnitude
is the same for each image.
Simulations of the left column are in a two-phase clumpy medium,
in which the clumps are spherical, $30$ times denser than the ICM,
and have a volume filling factor $\fc=0.1$.
Simulations of the right column are in a medium with
a fractal cloud of dimension $D=2.7$, filling factor $\fc=0.1$,
and densities having an exponential distribution (tending toward lognormal)
with an average that is $30$ times the ICM density,
and maximum that is $260$ times the ICM density.
In all cases the dust is characterized by a scattering albedo $\omega=0.6$
and an angular scattering phase function parameter
$ g = \langle \cos \theta \rangle = 0.6 $,
which are typical values for UV photons scattering off dust grains.

%analogous to a star or centrally condensed cluster of stars in an H II region.

Each row of images is at the same homogeneous optical depth \tauh,
which is the radial optical depth of absorption and scattering that would result
if the dust was distributed uniformly in the sphere instead of in clumps.
The upper row has $\tauh=1$ and lower row is for $\tauh=10$.
Increasing \tauh can be viewed as either increasing the dust abundance
or decreasing the wavelength of the photons, resulting in more absorption.
As \tauh increases the clumps become opaque,
creating the apparent shadows behind the clumps.
However scattering by the dust
causes photons to go behind the clumps and become absorbed,
thus diminishing the effect of what would otherwise be completely dark shadows
in the case of no scattering.
As the clumps become opaque absorption occurs more at the clump surfaces.

\begin{figure}
\label{fractal+mega_grain}                               
  \epsfysize=5.5in
  \centerline{\vbox{\epsfbox{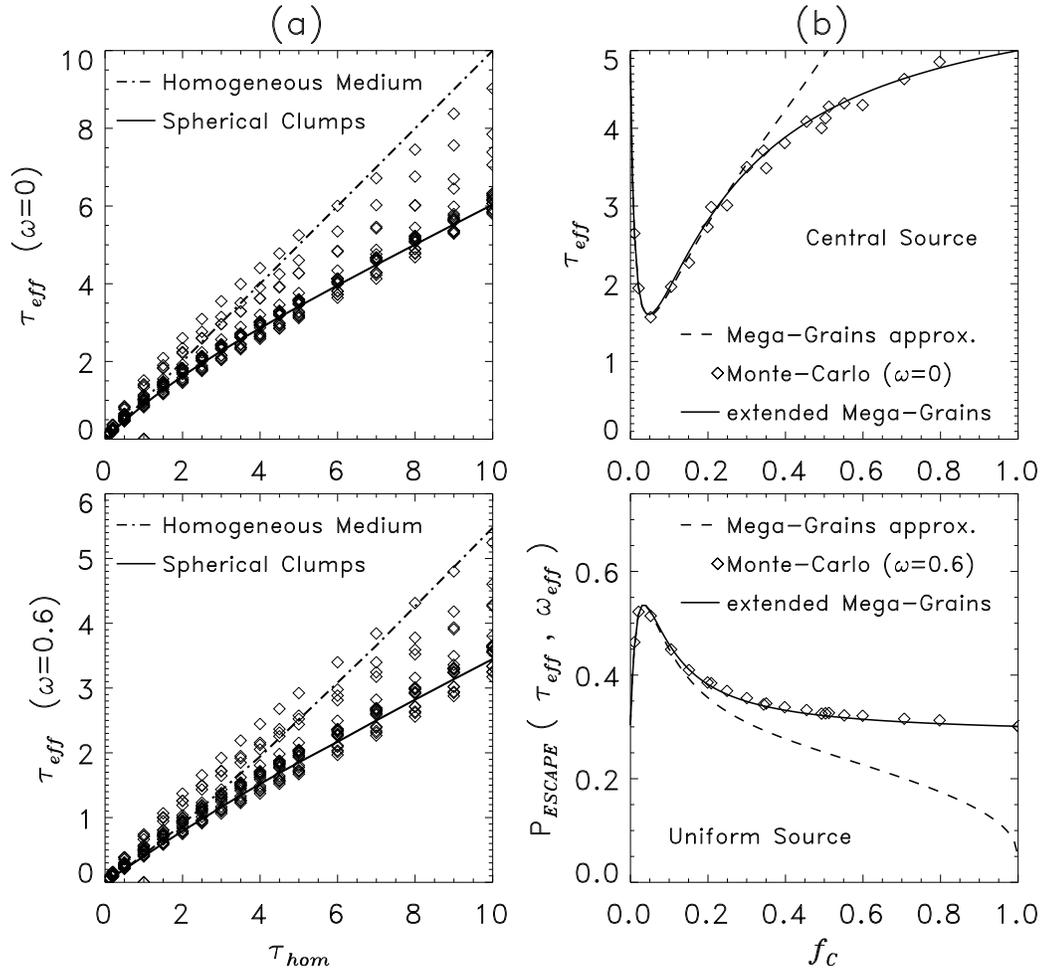}}}
\caption{
(a) Monte Carlo computations of
the effective optical depths of 20 random fractal dust clouds (diamonds),
plotted versus and compared with
equivalent homogeneous optical depth (dashed curve),
and effective optical depth of related two-phase clumpy medium (solid curve).
The upper graph is for no scattering and the lower graph is with scattering.
(b) Comparison of Monte Carlo simulations of radiative transfer
in two-phase clumpy media with Mega-Grains approximation. Horizontal
axis is the volume filling factor of spherical clumps,
upper graph is for central source and no scattering,
lower graph is for uniformly distributed emitters with scattering.
See text for further explanation. }
\end{figure}

By observing the fraction of photons that escape in
the case of a central source,
an effective optical depth is defined as
$ \taueff = -\ln( L_{escape} / L_{emit} ) $.
Usually \taueff $<$ \tauh in a clumpy
medium and the functional relationship is nonlinear.
This can be seen in Fig.\ref{fractal+mega_grain}(a),
were the effective optical depths of 20 randomly created fractal dust clouds
(with $D=2.5$ and $\fc=0.065$)
are plotted versus the equivalent homogeneous optical depth.
The upper graph is for no scattering and the lower graph is with scattering.
Also plotted for comparison are the homogeneous optical depth,
and effective optical depth of related two-phase clumpy medium (solid curve)
having the same \fc, and clump densities equal to the average density of
fractal clouds. Most of the fractal clouds have \taueff that varies
as a function of \tauh similar to the two-phase clumpy medium.
However, many of the fractal clouds modeled have \taueff that exceeds \tauh
for low values of \tauh. This is because the position of the source
relative to the fractal cloud is more important than in the
case of two-phase clumpy medium. The spherical clumps are small and
their distribution is uniformly random so the relative position of the
source is not so important, whereas the fractal cloud is a more
connected set, and even though \fc is small, if the source is
near to any part of the cloud, it is near to a lot of the cloud
and thus its emission is more likely to be absorbed or scattered.

\section{Analytic Approximations}

Since Monte Carlo simulations can require a large amount of computer time,
it is useful to have analytical approximations
for the basic results of radiative transfer: the fraction of photons escaping
and the fraction of photons absorbed in each phase of the medium.
In the case of a spherical homogeneous medium with
uniformly distributed emitters, the
escape probability formula of Osterbrock \cite{oster89}
is an exact solution when
there is absorption only.
\begin{equation}
\label{Oster_EP}                               
	P_0(\tau) = \frac{3}{4\tau} \left[ 1 - \frac{1}{2\tau^2} +
	\left( \frac{1}{\tau} + \frac{1}{2\tau^2} \right) e^{-2\tau} \right]
\end{equation}
Lucy et al. \cite{Lucy91} suggested a formula that extends any
absorption only escape probability to approximately
include the effects of scattering,
\begin{equation}
\label{Lucy_EP}                               
	P(\tau,\omega) = \frac{P_0(\tau)}{1-\omega[1-P_0(\tau)]}
\end{equation}
where $\omega>0$ is the scattering albedo,
the optical depth $\tau$ includes both absorption and scattering,
and $P_0$ is any escape probability for $\omega=0$.
Lucy's formula is based on the assumption that the scattered photons mimic
the photons emitted uniformly by the sources
so that the $\omega=0$ escape probability formula applies recursively.

The combination of Eqs.(\ref{Oster_EP}) and (\ref{Lucy_EP}), which we call the
Osterbrock-Lucy formulae, was tested extensively against
Monte Carlo radiative transfer simulations and was found to be a reasonable
approximation of the fraction of photons escaping from a homogeneous medium.
However, since the angular distribution of the scattered photons is ignored
in Lucy's approximation, the formula is exact
for only a single value of the angular scattering parameter
$ g = \langle \cos \theta \rangle $, where $\theta$ is the deflection angle,
and this value also depends $(\tau,\omega)$.
Coincidentaly, the $g$ dependence
of escape probability validity
follows the scattering properties of silicate and graphite dust,
that is, for low optical depths the escape probability agrees
with the isotropic scattering ($g=0$) case, and as optical depth increases
the agreement shifts toward more forward scattering cases ($g \rightarrow 1$).

For the case of a two-phase clumpy medium,
Hobson \& Padman \cite{hobpad93}
provide formulae approximating
the effective radiative transfer properties
by assuming spherical clumps and treating them as ``Mega-Grains''.
The upper graph in Fig.\ref{fractal+mega_grain}(b)
compares the effective radial optical depth of a spherical
clumpy medium obtained from the Mega-Grains approximation (dashed curve)
to \taueff derived from Monte Carlo simulations (diamonds),
over the full range of clump filling factor \fc,
in the case of $\tauh = 5.0$ and absorption only ($\omega=0$).
Each clump has a radius which is 5\% of the radius of the spherical region,
and density which is $100$ times the ICM density.
The graph shows that the Mega-Grains
approximation is valid for $ \fc < 0.25 $, but overpredicts \taueff
at larger filling factors.
By introducing another dependence on \fc,
the Mega-Grains approximation
can be extended to the full range of $ 0 \leq \fc \leq 1 $ (solid curve),
fitting the Monte Carlo results better and reproducing the correct
asymptotic value of \taueff = \tauh in the $\fc \rightarrow 1$ limit.

The Mega-Grains (MG) approximation gives the effective
optical depth \taueff and effective albedo \weff of the clumpy medium.
Using these parameters,
the escaping fraction of photons for the case of a uniform source
can be computed by substituting \taueff and \weff
directly into the Osterbrock-Lucy escape probability formulae,
Eqs.(\ref{Oster_EP}) and (\ref{Lucy_EP}).
The lower graph in Fig.\ref{fractal+mega_grain}(b)
compares this analytically computed escaping fraction
to the Monte Carlo simulations of uniformly distributed emitters,
including scattering.
The standard MG
approximation gives agreement with Monte Carlo only for $ \fc < 0.15 $.
Introducing another dependence on \fc in the MG formulae
(scaling the clump radius by $1-\fc$)
and using the escape probability formulae to get the effective albedo
of each clump,
improves the agreement with Monte Carlo results for the full range of \fc.
The escaping fractions determined by the combination of
the extended MG and escape probability approximations are
found to be in reasonable agreement
with  Monte Carlo results for $ 0 < \tauh \leq 40 $ and
$ 0 \leq \fc \leq 1 $. These approximations can be applied to the case
of a uniformly distributed source in a disk by using an effective radius:
$\Reff = 3Rh/(R + 2h)$,
where $R$ is the actual radius and $h$ is the half-thickness of the disk.

We have also formulated equations for estimating what fraction of
photons get absorbed in clumps and what fraction in the ICM,
for the cases of central and uniform source, and find the equations
to be reasonable approximations of the Monte Carlo results.
These absorbed fractions
are necessary for computing the dust temperatures and
the resulting infrared emission.
A test of the
approximations that needs to be performed is to check
how well the dust temperatures thus calculated match
the distribution of dust temperatures from the
corresponding Monte Carlo simulation.

\begin{figure}
\label{tau_wavelen}                               
  \epsfxsize=5.2in
  \centerline{\vbox{\epsfbox{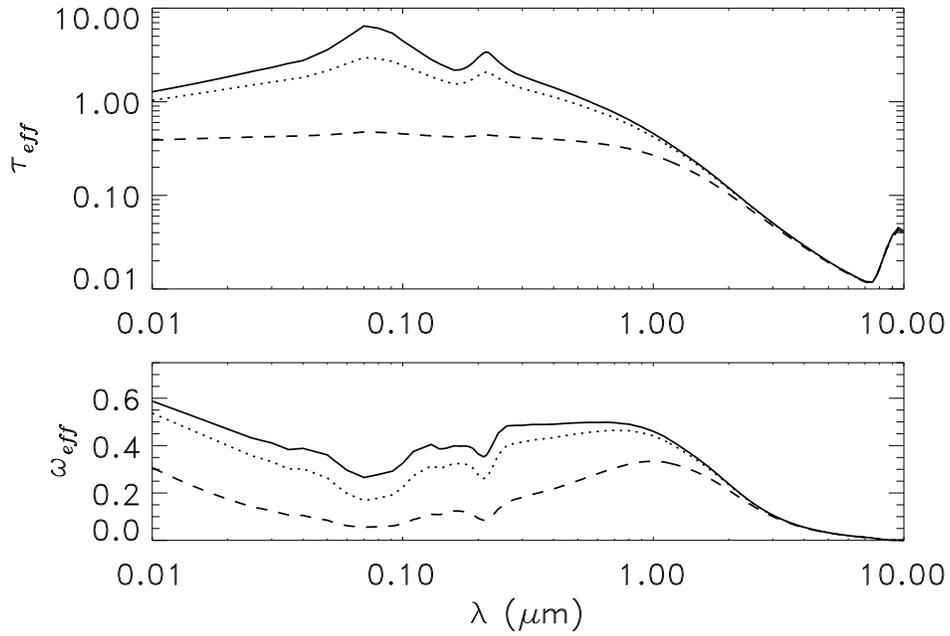}}}
\caption{
\small{
Radiative transfer properties of graphite and silicate dust
distributed in a sphere three different ways.
Solid curves are \tauh and albedo $\omega$ from the center of sphere
for the case of a homogeneous medium.
Dotted curves are \taueff and \weff for a two-phase clumpy medium
with $\fc=0.05$ and clump to ICM density ratio of $\rho_c / \rho_{icm} = 10^2$.
Dashed curves are for a clumpy medium
with $\fc=0.01$ and  $\rho_c / \rho_{icm} = 10^4$.
}}
\end{figure}

\section{Summary}

The degree of clumpiness of a medium
is as important as the total dust mass and scattering albedo
for the transfer of radiation.
For example, Fig.\ref{tau_wavelen}
compares the effective optical depth (absorption and scattering)
and effective albedo
of spherical regions of dust with different degrees of clumpiness,
computed using the MG approximation,
as a function of photon wavelength.
The dust is composed of equal amounts of graphite and silicates by mass,
having the $ a^{-3.5} $ grain size distribution for
$ 0.001 {\mu}m < a < 0.25 {\mu}m $.
The solid curves are the homogeneous case ($\fc=0$) of no clumps,
with dust mass density of $ 1.6 \times 10^{-23} \, {\textstyle {gm/cm^3}} $.
The dotted curves are for the case
$\fc = 0.05$ with $ \rho_c / \rho_{icm} = 100 $,
and the dashed curves are for
the extreme case of $\fc = 0.01$ with $ \rho_c / \rho_{icm} = 10^4 $.
Each clump has a radius of $0.01$ pc,
and the radius of the spherical region is $0.6$ pc.
All cases have the same total mass of dust.
From the graphs it is evident that
the effective radiative transfer properties of the dusty medium
can be radically affected by the degree of clumpiness.

Simulations of radiative transfer in a fractal distribution of dust indicate
that the medium becomes more transparent as the fractal dimension decreases.  Since the filling factor follows the fractal dimension,
this behavior is similar to
that of a two-phase clumpy medium, but there can be significant
qualitative and quantitative differences,
as seen in Figs.1 and \ref{fractal+mega_grain}(a).
However, it may be possible to use
the Mega-Grains approach to approximate the effective optical depth
of a random fractal cloud when the sources are not correlated with the cloud,
as shown by the solid curves in Fig.\ref{fractal+mega_grain}(a).
Those smooth solid curves are actually created by the MG approximation
for the case of spherical clumps with radii $R_c = 0.05$ of the medium radius
(MG agrees with Monte Carlo).
The parameters $\fc=0.065$ and $ \rho_c / \rho_{icm} = 16.2 $
used in the MG approximation match
the filling factor and average density of the fractal cloud,
but the radii of the clumps is a free parameter,
and the value $R_c = 0.05$ happens work in this case.
However, for the case shown in Fig.1, $R_c = 0.05$
does not work, as seen by the different values of \taueff resulting
when \tauh increases. So either an effective $R_c$ needs to be determined
as a function of cloud fractal dimension, or some other generalization
of the MG approximation is needed.


\clearpage

\begin{thebibliography}{99}

\bibitem{boisse90}  Boiss\'e, P. 1990, A\&A, 228, 483

\bibitem{elm97}  Elmegreen, B.G. 1997, ApJ, 477, 196

\bibitem{hs93}  Hobson, M.P. \& Scheuer, P.A.G. 1993, MNRAS, 264, 145
\bibitem{hobpad93}  Hobson, M.P. \& Padman, R. 1993, MNRAS, 264, 161

\bibitem{Lucy91}  Lucy, L.B., Danziger, I.J., Gouiffes, C., Bouchet, P. 1991,
	in Supernovae, ed. S.E. Woosley (New York: Springer-Verlag), p.82

\bibitem{LuxKob95}  Lux, I. \& Koblinger, L. 1995,
	Monte Carlo Particle Transport Methods:
	Neutron and Photon Calculations (Boca Raton, CRC Press), p.40

\bibitem{norman96}  Norman, C.A. \& Ferrara, A. 1996, ApJ, 467, 280

\bibitem{oster89}  Osterbrock, D.E. 1989, Astrophysics of Gaseous Nebulae and
   Active Galactic Nuclei (Mill Valley, CA: univ. Science Books), Appendix 2

\bibitem{witt77}  Witt, A.N. 1977, ApJSup, 35, 1

\bibitem{wittgor96}  Witt, A.N. \& Gordon, K.D. 1996, ApJ, 463, 681

\end{thebibliography}
\end{document}